\documentclass[
prd,%
preprint%
,secnumarabic%
,amssymb, amsmath,nobibnotes, aps]{revtex4}
\usepackage{epsfig}%
\usepackage{graphicx}%
\usepackage{multirow}
\expandafter\ifx\csname package@font\endcsname\relax\else
 \expandafter\expandafter
 \expandafter\usepackage
 \expandafter\expandafter
 \expandafter{\csname package@font\endcsname}%
\fi

\begin{document}

\title{ Why LHC?}

\author{ D.P.Roy }%
\affiliation{Homi Bhabha Centre for Science Education, Tata Institute of Fundamental Research,\\
V. N. Purav Marg, Mumbai-400088, India
}
\def\be{\begin{equation}}
\def\ee{\end{equation}}
\def\al{\alpha}
\def\bea{\begin{eqnarray}}
\def\eea{\end{eqnarray}}

\begin{abstract}
I discuss LHC physics in the historical perspective of the progress in particle physics. After a recap of the Standard Model of particle physics, I discuss the high energy colliders leading up to LHC and their role in the discovery of these SM particles. Then I discuss the two main physics issues of LHC, i.e. Higgs mechanism and Supersymmetry. I briefly touch upon Higgs and SUSY searches at LHC along with their cosmological implications.

\end{abstract}

\maketitle
\section{Basic Constituents of Matter and their Interactions (Standard Model)}

   As per the Standard Model the basic constituents of matter are a dozen of spin 1/2  particles called matter fermions. These are the three pairs of leptons - electron, muon and tau along with their associated neutrinos, and the three pairs of quarks - up, down, strange, charm, bottom and top. They are listed in Table 1 below along with their masses, shown in GeV units. The charged lepton masses span three orders of magnitude; and the same holds for the (constituent) quark masses. The neutrino masses are negligibly small in comparison.

   \begin{table}[h!]
	   \centering
	   \begin{tabular}{|c|c c c|}\hline
		    \multirow{2}{*}{Leptons} & $\nu_e$ & $\nu_\mu$ & $\nu_\tau$ \\
		    & $e(0.0005)$ & $\mu(0.1)$ & $\tau(1.8)$ \\\hline
		    \multirow{2}{*}{Quarks} & u(0.3) & c(1.5) & t(175) \\
		    & d(0.3) & s(0.5) & b(5) \\\hline
	   \end{tabular}
	   \caption{The matter fermions shown along with their masses (in GeV).}
	   \label{tab1}
   \end{table}


   Each pair differ by 1 unit of electric charge - charge 0 and -1 for neutrinos and charged leptons and 2/3 and -1/3 for the upper and lower quarks. This simply reflects the fact that each pair is a weak isospin doublet with $I_3 = \pm 1/2$. Apart from this electric charge the quarks also carry a new kind of charge called colour charge, which is responsible for their strong interaction that binds them together inside nuclear particles (hadrons). Thus, leaving aside gravity, which is too weak to have any perceptible effect, there are three basic interactions between these particles - i.e. strong, weak and electromagnetic interactions. They are all gauge interactions, which are completely determined by the respective gauge groups - SU(3), SU(2) and U(1). In particular, they are all mediated by the exchange of spin 1 (vector) particles called gauge bosons, whose couplings are proportional to the respective gauge charges.
   Table 2 summarizes the three gauge interactions along with their gauge groups, gauge bosons and gauge charges. It also shows the scattering diagrams (Feynman diagrams) and scattering amplitudes corresponding to these three interactions.

   \begin{table}
	   \centering
	   \begin{tabular}{|c|c|c|c|c|c|}\hline
		   Interaction & Gauge group & Gauge boson & Gauge charge &
		   Scattering diagram & Scattering amplitude \\\hline
		   \raisebox{7ex}{Strong} & \raisebox{7ex}{SU(3)} & \raisebox{7ex}{g}
		   & \raisebox{7ex}{C} & \includegraphics[scale=.75]{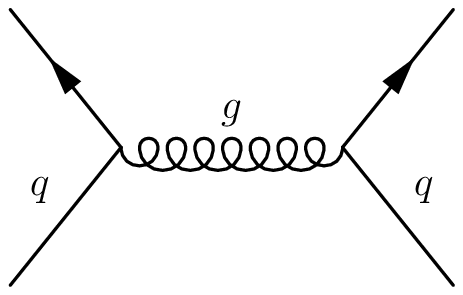} &
		   \raisebox{7ex}{$\alpha_s C_q^2/Q^2$}\\\hline
		   \raisebox{7ex}{EM} & \raisebox{7ex}{U(1)} & \raisebox{7ex}{$\gamma$}
		   & \raisebox{7ex}{e} & \includegraphics[scale=.75]{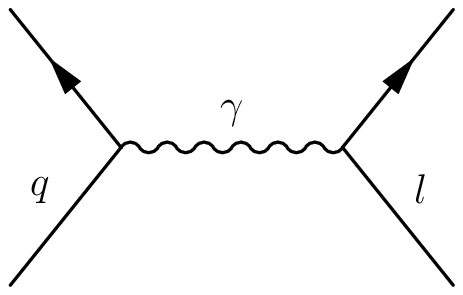} &
		   \raisebox{7ex}{$\alpha e_q\,e_l/Q^2$}\\\hline
		   \raisebox{7ex}{Weak} & \raisebox{7ex}{SU(2)} & \raisebox{7ex}{W,Z}
		   & \raisebox{7ex}{I} & \includegraphics[scale=.8]{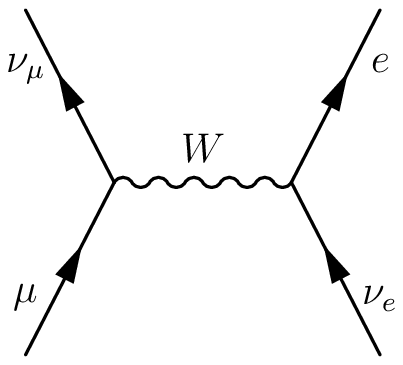} &
		   \raisebox{7ex}{$\alpha_W/(Q^2-M_W^2)$}\\\hline
	   \end{tabular}
	   \caption{Summary of the three gauge interactions.}
	   \label{tab2}
   \end{table}

   The electromagnetic (EM) interaction between quarks and charged leptons is mediated by the massless gauge boson (photon), whose coupling strength is proportional to their electric charges. The constant of proportionality is the square root of the fine structure constant $\sqrt \alpha$. Thus the scattering amplitude is given by the product of the two coupling strengths, representing the photon emission and absorption amplitudes, and the photon propagation amplitude (propagator) $1/Q^2$ ,  where $Q^2$ is the Lorentz invariant four-momentum square transferred between the particles. Likewise, the strong interaction between quarks is mediated by a massless gauge boson called gluon, whose coupling strength is proportional to their colour charge. The constant of proportionality is called $\alpha_s$, in analogy with the fine-structure constant $\alpha$. And the theory of strong interaction is called quantum chromodynamics (QCD) in analogy with quantum electrodynamics (QED). There is however one important difference between the two interactions, which arises from the nonabelian nature of the gauge group SU(3). This implies that unlike the electric charge the colour charge is a vector quantity, which can take three possible direction in an abstract space called red, blue and green. Of course the colour charges of the three constituent quarks of a nucleon cancel out vectorially, so that the nuclear particles do not carry any net colour charge just like the atoms do not carry any net electric charge. Because of the vector nature of the colour charge, however, the gluon has a colour charge and hence self interaction, while the photon has no electric charge and so no self interaction. The gluon self interaction leads to the confinement phenomenon, which means the quarks are perpetually confined inside nuclear particles.

   The weak interaction is mediated by massive vector bosons, W$^\pm$ and Z$^0$. The charged W boson couples to each pair of leptons and quarks with a universal coupling strength $\sqrt{\alpha_W}$, because they all carry the same weak charge, I = 1/2. But the W boson mass term appears in its propagator and the resulting weak scattering amplitude shown in Table 2. This means that the weak interaction has a short range, corresponding to the inverse of $M_W$. Since the emission of a W boson implies the creation of its rest mass energy, the range is $\sim M_W^{-1}$ in natural units by the uncertainty principle. Note that in quantum field theory the absorption of a particle is equivalent the emission of its antiparticle. Thus the weak scattering amplitude of Table 2 describes not only the scattering process $\nu_e \mu \rightarrow e \nu_\mu$, but also the corresponding decay process.

\be
\mu\rightarrow \nu_\mu e \bar \nu_e                                                                                                                       \ee
Moreover, for a low energy process like muon decay we have $Q^2 << M_W^2$, so that the muon decay amplitude reduces to
\be
   \alpha_W/M_W^2 = (\sqrt{2}/\pi)G_F
\ee
where $G_F \simeq 1.17 \times 10^{-5} GeV^{-2}$ is the Fermi coupling constant.

   The weak and electromagnetic interactions have been unified into a SU(2)$\times$U(1) electroweak theory. This means that their coupling constants are related to one another, i.e.
   \be
\alpha_W=\alpha/Sin^2 \theta_W \simeq \alpha/0.23                                                                                        \ee
where $\theta_W$ is the mixing angle  between the neutral gauge bosons of the two groups. So the weak interaction is intrinsically as strong as the electromagnetic. The reason the low energy weak processes like muon decay are so weak is their short range, as reflected in the $M_W^2$ term in the propagator. An immediate prediction of this theory is the W boson mass. From eqs. (2), (3) and the measured value of $G_F$ from muon decay rate one gets
\be
 M_W \simeq 80 GeV, \quad M_Z =M_W/Cos \theta_W \simeq 91 GeV
\ee
This completes the summary of matter fermions and gauge bosons along with their masses. Let us see where are they?

   The up and down quarks are the constituents of proton and neutron. So together with the electron they constitute all the visible matter of the universe. All the heavier quarks and leptons decay weakly into the lighter ones, just like the muon decay process seen above. So they do not occur freely in nature. But they can be produced in cosmic ray or laboratory experiments. The muon was discovered in cosmic ray experiment in the mid forties. So was the strange quark in the form of K meson. The neutrinos are stable but very hard to detect because they have only weak interaction with matter. The $\nu_e$ was discovered in an atomic reactor experiment in the mid fifties [1]. This was followed by the discovery of $\nu_\mu$ in a proton accelerator experiment at BNL [2]. The first cosmic ray observation of neutrino ($\nu_\mu$) came from the Kolar Gold Field experiment [3]. Finally, the $\nu_\tau$ was discovered in an emulsion experiment at the fixed target proton accelerator at Fermilab [4]. The rest have come mainly from the electron-positron and proton-antiproton colliders over the past 30-40 years. First came the electron-positron colliders in the seventies, leading to a windfall of discoveries - charm quark, tau lepton, bottom quark and the gluon. This was followed by the discovery of W and Z bosons in 1983 and top quark in 1995 at the proton-antiproton colliders of CERN and Fermilab respectively. The colliders have been the main workhorse of high energy physics over the past thirty odd years and are likely to remain so over the next thirty odd years. So let us discuss them briefly.

\section{High Energy Colliders}

\begin{figure}[h!]
	\begin{center}
		\includegraphics[width=.5\textwidth,height=4cm]{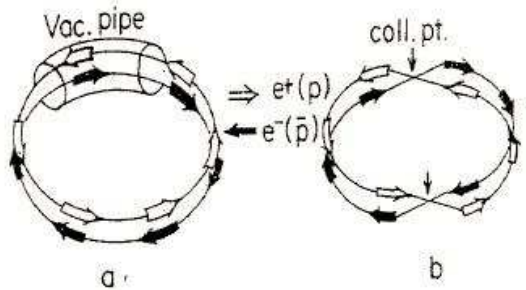}
	\end{center}
	\caption{Schematic diagram of $e^+ e^-$ (or $\bar p p$) collider in (a) acceleration and (b) collision mode.}
	\label{fig1}
\end{figure}

   An $e^+ e^-$ or $\bar p p$ collider is a synchrotron machine, where the particle and antiparticle beams are accelerated inside the same vacuum pipe (Fig 1) using the same set of bending magnets and accelerating cavities (not shown). Thanks to their equal mass and opposite charge the two beams go round in identical orbits on top of one another throughout the course of acceleration. When the acceleration is complete they are made to collide head on by flipping their orbits with a magnetic switch. In this mode the beams keep on colliding repeatedly at the collision points, which are surrounded by detectors to see what comes out of the collisions. The main advantage of a collider over a fixed target machine is an enormous gain in the centre of momentum frame (CM) energy, which is the effective energy available for particle production, at very little extra cost. For a collider, the squared CM energy $s$ is simply related to the beam energy $E$ by $s = (2E)^2$, while for a fixed target machine with target mass $m$ it is given by $s = 2mE$. Thus a collider beam energy $E$ is equivalent to a fixed target machine beam energy $ E^\prime$, where
   \be
E^\prime=2 E^2/m
\ee
The Tevatron $\bar p p$  collider at Fermilab has a beam energy $E = 1$ TeV. Thus with a proton target mass $m \simeq 1$ GeV, this is equivalent to a fixed target machine beam energy of 2000 TeV. For an $e^+ e^-$ collider the energy gain is another factor of 2000 higher, because the target mass is a factor of 2000 lower.

   Let us compare the relative merits of the $\bar p p$  and $e^+ e^-$ colliders. The proton (antiproton) is nothing but a bunch of quarks (antiquarks) and gluons. Thus the $\bar p p$  collision can be used to study  $\bar q q$ interaction, which can probe the same physics as the $e^+ e^-$ interaction (Fig 2).

   \begin{figure}[h!]
	   \begin{center}
		   \includegraphics[width=.7\textwidth,height=4cm]{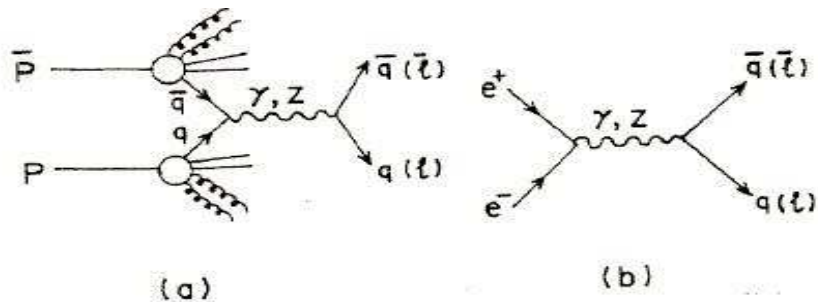}
	   \end{center}
	   \caption{Comparison of $\bar p p$ and $e^+ e^-$ collisions.}
	   \label{fig2}
   \end{figure}

   Of course, one has to pay a price in terms of the beam energy, since a quark carries only about 1/6th of the proton energy$^1$ \footnotetext[1]{The proton energy momentum is shared about equally between quarks and gluons; and since there are 3 quarks, each one has a share of $\sim 1/6$ on the average}. Thus
\be
   s^{1/2}_{e^+ e^-} \simeq    s^{1/2}_{q \bar q } \simeq (1/6)s^{1/2}_{\bar p p}
\ee
i.e. the energy of the $\bar p p$ collider must be 6 times larger than that of the $e^+ e^-$ collider to give the same interaction energy. This is a small price to pay, however, considering the immensely higher synchrotron radiation loss for the $e^+ e^-$ collider. The amount of synchrotron radiation loss per turn is
\be
   \Delta E \simeq \frac{4 \pi}{3} \frac{e^2 E^4}{m^4 \rho}
\ee
where e, m, and E are the particle charge, mass and energy and $\rho$ is the radius of the
ring. This quantity is much larger for the $e^+ e^-$ collider due to the small electron
mass, which would mean a colossal energy loss and radiation damage. These are reduced by
increasing the radius of the ring, resulting in a higher construction cost. The point is
best illustrated by a comparison between the CERN $\bar p p$ collider and its large $e^+ e^-$ (LEP) collider, which were both built to produce W and Z bosons. The CM energy of LEP-I was 100 GeV against the 6 times higher CM energy of the  $\bar p p$ collider. But the radius of LEP was 5 km and cost 1 billion dollars against the radius of only 1 km and cost of 300 million dollars for the $\bar p p$ collider. In fact, out of the total construction cost of 300 million dollars, 200 million went into building the fixed target machine (SPS) and only about a 100 million in converting it into a $\bar p p$  collider. Thus the $\bar p p$ collider has an obvious cost advantage over the $e^+ e^-$ machine.

   On the other hand, the $e^+ e^-$ collider has a great advantage over $\bar p p$ as a precision machine. This is because one can tune the $e^+ e^-$ energy to the desired particle mass ($M_Z$), which cannot evidently be done with the quark-antiquark energy. Thus one could produce about a million of $Z \rightarrow e^+ e^-$ events per year at LEP against about only a few dozen at the CERN $\bar p p$ collider. Moreover, the $e^+ e^-$ collider signals are far cleaner than the $\bar p p$ collider, since one has to contend with the debris from the spectator quarks and gluons in the latter case. In summary, the $\bar p p$ collider is more suited for surveying a new energy domain because of its cost advantage, while the $e^+ e^-$ is better suited as a precision machine for a detailed follow-up study. This principle has guided the history of high energy colliders, as we see from the Table 3.

   \begin{table}
	   \centering
	   \begin{tabular}{|c|c|c|c|c|c|c|}\hline
	    \multirow{5}{*}{70's} & SPEAR & Stanford & \multirow{5}{*}{$e^+e^-$} & $3+3$ & & charm,$\tau$\\
	    & DORIS & Hamburg  & & $5+5$   &         & bottom\\
	    & CESR  & Cornell  & & $8+8$   & $125$ m & bottom\\
	    & PEP   & Stanford & & $18+18$ &         &       \\
	    & PETRA & Hamburg  & & $22+22$ & $300$ m & gluon \\\hline
	    \multirow{2}{*}{80's} & TRISTAN & Japan & $e^+e^-$ & $30+30$ & & \\
	    & S$\overline{\mbox{P}}$PS & CERN & $\bar pp$ & $300+300$ & 1 km & W,Z boson\\\hline
	    \multirow{5}{*}{90's} & TEVATRON & Fermilab & $\bar pp$ & $1000+1000$ & \multirow{5}{*}{5 km} & Top\\
	    & SLC      & Stanford & $e^+e^-$ & $50+50$   & & Z\\
	    & LEP-I    & CERN     &          & $50+50$   & & Z\\
            & (LEP-II) &          &          & $100+100$ & & W\\
	    & HERA     & Hamburg  & $ep$     & $30+800$  & &  \\\hline
	    2009 & LHC & CERN & $pp$ & $7000+7000$ & 5km & Higgs, SUSY\\\hline
	    2??? & ILC & ??? & $e^+e^-$ & $500+500$ & & \\\hline
	   \end{tabular}
	   \caption{Past, present and proposed colliders.}
	   \label{tab3}
   \end{table}

   First came Stanford $e^+ e^-$ collider (SPEAR) with a CM energy of 6 GeV, which was
   responsible for the discoveries of charm quark [5] and $\tau$ lepton [6]. This was
   followed by DORIS at Hamburg and CESR at Cornell with CM energies of $\sim 10$ GeV,
   which were dedicated to the study of bottom quark. The CM energies of the above $e^+
   e^-$  colliders were similar to those of the fixed target proton synchrotrons at BNL
   and Fermilab. In fact the charm quark was simultaneously discovered at BNL PS [7]; and
   the first evidence of the bottom quark came from the Fermilab PS [8]. But the bulk of
   our knowledge of the charm and bottom quarks has come from these $e^+ e^-$ colliders.
   Next came the PEP at Stanford and PETRA at Hamburg, raising the CM energy to $\sim 40$
   GeV; which resulted in the discovery of gluon at PETRA [9]. Then came the CERN $\bar p
   p$  collider with a CM energy of $\sim 600$ GeV, corresponding to an interaction energy of 100 GeV. This was enough for the discovery of W and Z bosons [10]. It was followed by the Fermilab $\bar p p $ collider (Tevatron) with a CM energy of $\sim 2$ TeV, corresponding to an interaction energy of $\sim 350$ GeV. This was adequate for the pair production of top quark [11]. Then came the $e^+ e^-$ colliders with enhanced CM energy of 100 GeV at Stanford (SLC) and CERN (LEP-I) for a precision study of the Z boson properties. The latter was upgraded to LEP-II, with a CM energy of $\sim 200$ GeV, for pair production of W± bosons and precision study of their properties. Next came the ep collider (HERA) at Hamburg, which was dedicated to a detailed study of the quark substructure of proton. Finally, the large hadron collider (LHC) is a $pp$ collider, which has recently started operation in the LEP tunnel at CERN. The advantage of having two colliding proton beams is to achieve a higher luminosity than a $\bar p p$ collider. The scheduled CM energy is 14 TeV, corresponding to an interaction energy of a little over 2 TeV. This will extend the search of new particles into the TeV range. Hopefully this will be followed by an $e^+ e^-$ collider in future called international linear collider (ILC) for a precision study of this energy range. It has to be a linear collider because the synchrotron radiation is much too large for a circular $e^+ e^-$ collider to operate at this energy range. And it has to be international because no single country or continent can afford its cost! Of course one may ask here that, since we have seen all the basic constituents of matter and the carriers of their interactions, why need we carry forward the new particle search into the TeV range? The answer is that the story is not complete yet. As we shall see in the next two sections, we expect several new particles in this mass range in the form of Higgs boson(s) and supersymmetric particles, which will hopefully be discovered at LHC.

   Let us close this section with a brief discussion of the collider signatures of the abovementioned matter fermions and gauge bosons. The bottom and charm quarks as well as the $\tau$ leptons are pair produced in an   $e^+ e^-$ collider via a virtual photon,
\be
 e^+ e^- \overset{\gamma}{\rightarrow} \bar c  c, \bar b  b, \tau^+ \tau^-
\ee
The outgoing pair of new particles comes out back to back because of momentum conservation. Their life-times are typically $ \sim$ picosecond, corresponding to decay ranges $c \tau \sim $  a few hundred microns. Thanks to the developments in silicon microchips technology, this is enough to tag these particles now by tracing their paths before decay. The gluon comes out via QCD radiation from one of the ordinary light quark pair produced in an $e^+ e^-$ collider, i.e.
\be
 e^+ e^- \overset{\gamma}{\rightarrow} \bar q  q g
\ee
where the rate and the energy and angular distributions of the three particles in the final state are predicted by QCD. By comparing these predictions with the observed rate and distributions of three-body final state one infers the third object to be a gluon. One may wonder here, how do the quarks and gluon escape from the confining force of QCD, mentioned earlier. This is possible because in quantum mechanics the vacuum is not empty. Instead it is full of particles like quarks and gluons, thanks to the uncertainty principle, which can account for their rest mass and kinetic energy. So the produced quarks and gluon can pick up extra quarks and gluons from the vacuum to come out as three colourless clusters of hadrons, carrying the respective momenta of the three produced particles. Each cluster has only a very small momentum spread coming from the intrinsic momenta of the vacuum particles, which are restricted by uncertainty principle to the inverse of the hadronic dimension, i.e. $< 1 {\rm fm}^{-1} (\sim 0.2 {\rm GeV})$. Thus each of the produced quarks and gluon emerge as a thin jet of hadrons.

   For W and Z boson production at a $\bar p p$ collider,
\be
 \bar q^\prime  q \overset{W^-}{\rightarrow} e^- \bar \nu_e,  \mu^- \bar \nu_\mu  \quad \&\quad        \bar q q \overset{Z}{\rightarrow} e^- e^+, \mu^- \mu^+
\ee
the kinetic energy of the incoming quark-antiquark pair is instantly converted into the rest mass energy of the W(Z) boson. So the direction of incoming momenta is lost; and the outgoing decay lepton pair often comes out nearly in a transverse direction to the beam pipe. Thus one gets a distinctive Z boson signal as a hard lepton-antilepton pair, each carrying a large transverse momentum $p_T  \simeq M_Z/2 \simeq 45$ GeV. The corresponding signal for W boson is a hard lepton with $p_T \simeq M_W/2 \simeq 40$ GeV, and an apparent imbalance of $p_T$ , since the neutrino escapes the detector without a trace. This $p_T$ imbalance (or missing-$p_T$ ) serves as a signature for the escaping neutrino. The top signal is more complex as it involves pair production of top, followed by their leptonic and hadronic decays, i.e.
\be
\bar q q \rightarrow \bar t t \rightarrow \bar b b W^+ W^- \rightarrow \bar b b \bar q q l \nu
\ee
One requires leptonic decay of one top to provide the trigger and hadronic decay of the other for reconstructing their masses. The large kinetic energy released in the top-antitop decays implies that the decay quarks and leptons are all hard (carry large $p_T$) and come out wide apart from one another. Thus one expects to see a hard isolated lepton along with a large number of hard jets and a missing-$p_T$. One can use the lepton isolation criterion along with the number and hardness of jets to extract the top signal from the SM background [12].

\section{Mass Problem - Higgs Mechanism }

   The mass terms of the weak gauge bosons would break the gauge symmetry of the Lagrangian, which is required for a renormalizable field theory. This is the most serious problem of the SM. In order to appreciate it consider the weak interaction Lagrangian of a scalar field $\phi$, which is a weak isospinor (doublet) carrying isospin I = 1/2; i.e.
\be
 L=(\partial_\mu \phi + i g \frac {\vec \tau}{2} \cdot \vec W_\mu \phi)^{\dagger}(\partial_\mu \phi + i g \frac {\vec \tau}{2} \cdot \vec W_\mu \phi)-\left[\mu^2 \phi^\dagger \phi + \lambda (\phi^\dagger \phi)^2\right]-(1/4) \vec W_{\mu \nu} \cdot \vec W_{\mu \nu}
\ee
where
\be
        \vec W_{\mu \nu}=\partial_\mu \vec W_\nu -   \partial_\nu \vec W_\mu -g \vec W_\mu \times \vec W_\nu
\ee
is the field tensor for the weak gauge boson $\vec W_\mu$ . Because of  the nonabelian nature of the gauge group SU(2) the gauge charge (isospin) is a vector quantity, represented by the three Pauli matrices, with the coupling constant $g^2 = 4 \pi \alpha_W$. Likewise the charged and the neutral W bosons form an isovector $ \vec W_\mu$. This is responsible for the last term in the field tensor (13), which leads to gauge boson self-interaction. Correspondingly, the gauge transformation on $\vec W_\mu  $ has an extra term compared to the EM case, i.e.
\be
 \phi \rightarrow e^{i \vec \alpha \cdot \vec \tau} \phi, \quad \vec W_\mu \rightarrow \vec W_\mu -(1/g) \partial_\mu \vec \alpha -\vec \alpha \times \vec W_\mu
\ee
This ensures gauge invariance of $ \vec W_{\mu \nu}^2$ , i.e. the last term of the Lagrangian, representing gauge kinetic energy and self-interaction. Evidently, the middle term, representing the scalar mass and self-interaction, is invariant under gauge transformation on $\phi$. Finally, the first term, representing scalar kinetic energy and gauge interaction, can be easily shown to be invariant under the simultaneous gauge transformations (14). But the addition of a gauge boson mass term,
\be
  -M^2 \vec W_\mu \cdot \vec W_\mu
\ee
would clearly break the gauge invariance of the Lagrangian.

   So the question is how to give mass to the weak gauge bosons without breaking the gauge invariance of the Lagrangian. The answer is provided by the celebrated Higgs mechanism of spontaneous symmetry breaking [13, 14]. It is based on the observation that the scalar mass term, $\mu^2 \phi^\dagger \phi$ , remains gauge invariant even for negative $\mu^2$ (imaginary mass). This is exploited to give mass to the gauge bosons through back door, as we see below. Fig 3 shows the scalar potential $\mu^2 \phi^\dagger \phi + \lambda (\phi^\dagger \phi)^2$ ,  as a function of the neutral component of the complex scalar field $\phi^0$ for (a) negative $\mu^2$ and (b) positive $\mu^2$. For the normal case of positive $\mu^2$ the minimum of the potential occurs at the origin, $\phi^0 = 0$. For the negative $\mu^2$, however, the minimum moves to a finite value of the field, i.e.
\be
  v=\sqrt{-\mu^2/\lambda}
\ee
The minimum of the potential corresponds to the ground state of energy (vacuum). Therefore, the vacuum corresponds to a finite value of the field $v$ (16), called its vacuum expectation value (vev).

\begin{figure}[h!]
	\begin{center}
		\includegraphics[width=.7\textwidth,height=4cm]{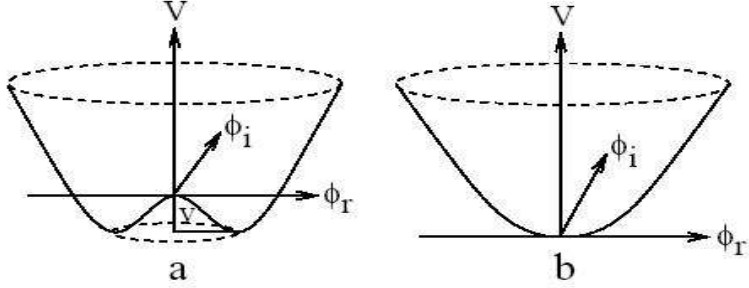}
	\end{center}
	\caption{The scalar potential, $V=\mu^2\phi^\dagger\phi+\lambda(\phi^\dagger\phi)^2$, plotted as a function of the complex scalar field $\phi$, for (a) negative $\mu^2$ and (b) positive $\mu^2$.}
	\label{fig3}
\end{figure}

Since the perturbative expansion in quantum field theory is stable only around a local minimum of energy, one has to translate the scalar field by the constant v,
\be
  \phi^0=v + h^0                                                                                                               \ee
Thus one gets a valid perturbative field theory in terms of the redefined field h. This represents the physical Higgs boson, while the three other components of the complex doublet field are absorbed to give mass and hence longitudinal components to the three gauge bosons.

   Substituting (17) in the first term of the Lagrangian (12) leads to a quadratic mass term for W boson with
\be
  M_W=\frac{1}{2} g v                                                                                                                  \ee
It also leads to a hWW coupling,
\be
  \frac{1}{2}g^2 v= g M_W                                                                                                             \ee
i.e. the Higgs coupling to gauge boson is proportional to the gauge boson mass. Similarly, its Yukawa coupling to quarks and leptons can be shown to be proportional to their respective masses, i.e.
\be
 y_{l,q}=\frac{m_{l,q}}{v}=\frac{1}{2} \frac{g m_{l,q}}{M_W}                                                                                         \ee
Indeed, this Yukawa coupling is the source of the quark and lepton masses in the SM. Finally, substituting (17) in the middle term of the Lagrangian (12) leads to a real mass for the physical Higgs boson,
\be
 M_h=v \sqrt{2 \lambda}=M_W(2 \frac{\sqrt{2 \lambda}}{g})                                                                                     \ee
   Substituting $M_W = 80$ GeV and $g = 0.65$ along with a perturbative limit on the scalar self-coupling $\lambda$, one gets an upper limit of a few hundred GeV for the Higgs boson mass. Therefore one hopes to discover this particle at the LHC. Moreover, we saw in eqs. (19) and (20) above that the Higgs couplings to quarks, leptons and gauge bosons are proportional to their respective masses. Thus the most important channels for Higgs search are its decay into heavy particle pairs,
\be
  h\rightarrow W^+ W^-, Z Z, \bar t  t,\bar b b, \tau^+ \tau^-                                                                                       \ee
For Higgs mass below the WW threshold, its dominant decay mode into the $b \bar b$ channel suffers from a large QCD background. In this region its radiative decay via W loop, $h\rightarrow \gamma \gamma$, offers a viable signature for Higgs search along with the $h\rightarrow \tau \tau$  decay.

   In the minimal supersymmetric extension of the standard model (MSSM), discussed in the next section, we need two Higgs doublets to give mass to the upper and lower fermions. Two complex doublets correspond to eight degrees of freedom, three of which are absorbed to give mass to the three weak gauge bosons. This leaves five physical Higgs bosons - two neutral scalars, a pseudo-scalar and a pair of charged Higgs bosons (h, H, A and H$^\pm$). While the prominent decay channels of the neutral scalars are same as eq. (22), those of the pseudo-scalar and the charged Higgs bosons are
\bea
A \rightarrow \bar t t, \bar b b, \tau^+ \tau^- \nonumber\\
H^+ \rightarrow t \bar b, \tau^+ \nu
\eea
While the Higgs decays into top and bottom quark channels suffer from large QCD backgrounds, the $\tau$ channels are more promising for both charged and neutral Higgs search. Moreover, one can use the distinctive $\tau$ polarization to enhance the signal/background ratio, particularly for the charged Higgs case [15]. A detailed account of both SM and MSSM Higgs boson searches can be found e.g. in ref [16].

   Before leaving this section, let us take a closer look at the basic principles underlying the Higgs mechanism. For the normal case of positive $\mu^2$ the ground state corresponds to $\phi = 0$, which is evidently invariant under gauge transformation, i.e the phase rotation of eq. (14). But for negative $ \mu^2$ the ground state moves to a finite value of $\phi$, which does not remain invariant under this phase rotation, i.e. the gauge symmetry is broken in the ground state (vacuum) because of the presence of a nonzero Higgs field of magnitude $v$. This symmetry breaking is responsible for giving mass to the weak gauge bosons. In fact one can clearly see from substituting (17) in the first term of the Lagrangian (12) that the gauge boson mass (18) arises from its interaction with this background Higgs field present in the vacuum. Likewise the quarks and leptons get their masses from their Yukawa interactions with this background Higgs field $v$, as shown in eq. (20). At the same time the Lagrangian (12) remains gauge invariant, as we have explicitly seen above. This is why the renormalization theory remains valid. This phenomenon of a symmetry being respected by the Lagrangian but broken by the ground state of a system is called spontaneous symmetry breaking (SSB). Of course it is nontrivial to show that the renormalization theory goes through in the presence of SSB. In fact t' Hooft and Veltman got the Nobel prize for precisely proving this.

   The phenomenon of SSB may appear as a mathematical artifice on first encounter. But it should be noted that examples of SSB abound in many braches of physics, from condensed matter to cosmology. The simplest example of SSB is a ferromagnet. At a high temperature its atomic spins are randomly oriented; and this disordered system has rotational symmetry because there is no preferred direction of spin. When cooled below a critical temperature, its atomic spins get aligned with one another in one direction as this ordered state corresponds to the ground state of energy. Thus the rotational symmetry of the EM interaction Lagrangian is broken by the ground state. We have a similar situation in the Higgs mechanism, except that the rotational symmetry is in the phase rather than the ordinary space. As per our present understanding of cosmology, such a phase transition occurred when the universe was a few picoseconds old and its temperature (i.e. the average kinetic energy of the constituent particles) a few hundred GeV. One can see from the finite temperature field theory how the effective mass of a particle changes with the medium, and in particular with the kinetic energy of the particles in the medium. It is at this point that the scalar mass square ($\mu^2$) became negative, giving mass to quarks, leptons and gauge bosons (Fig. 4). They readily decayed into the lightest quarks and leptons - u, d, and e. The next phase-transition occurred when the universe was about a microsecond old and its temperature around a GeV. This is when the u and d quarks coalesced to form protons and neutrons; and the universe became colour neutral. Next, when the universe was about a few seconds old and its temperature about a MeV, the neutrons were absorbed to form $\alpha$ particles. Finally, when the universe was about half a million years old and its temperature about an eV, the electrons were absorbed by protons and $\alpha$ particles to form neutral H and He atoms. At this stage the universe became transparent as matter decoupled from EM radiation. After this it started experiencing the small gravitational perturbations, leading to galaxy formation. The present age of the universe is about 15 billion years and temperature 1 meV (3 degrees Kelvin). But the story is not complete yet because of the hierarchy problem, which shall be discussed next.

\begin{figure}[h!]
	\begin{center}
		\includegraphics{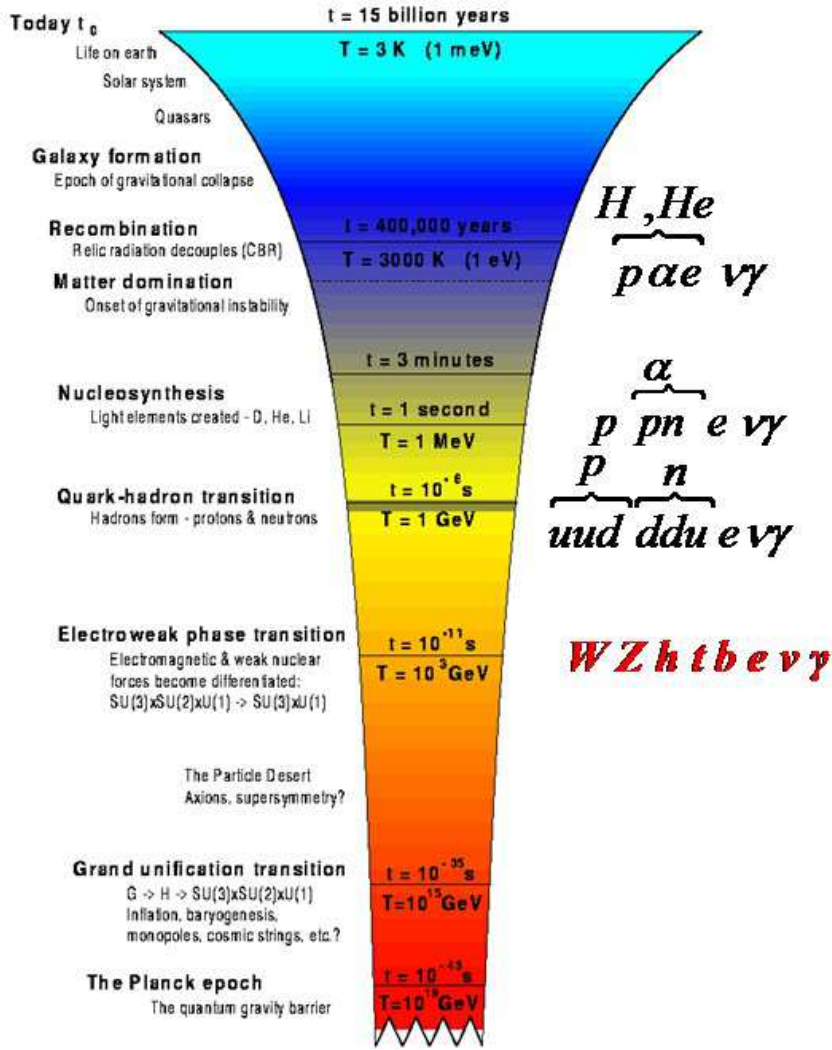}
	\end{center}
	\caption{Cosmological evolution of the universe - a brief history.}
	\label{fig4}
\end{figure}


\section{Hierarchy Problem - Supersymmetry }

   As we have seen above, giving mass to the weak gauge bosons via Higgs mechanism implies that the physical Higgs scalar mass also lies in the range of a few hundred GeV. This is problematic, however, because the scalar masses are known to get quadratically divergent quantum corrections. This comes from its interaction with other particles present in the quantum mechanical vacuum. For instance, its interaction with the other Higgs bosons present in the vacuum via the quartic self-interaction term of eq. (12) is illustrated in the Feynman diagram of Fig 5.

   \begin{figure}[h!]
	   \begin{center}
		   \includegraphics{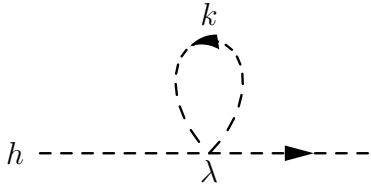}
	   \end{center}
	   \caption{The quadratically divergent quantum correction to the Higgs mass coming from its quartic self-interaction loop.}
	   \label{fig5}
   \end{figure}

Integrating over the momentum of the propagator loop results in a quadratically divergent contribution,
\be
   \lambda \int_{0}^{\Lambda} \frac{k^3 dk}{k^2-M^2} \propto [k^2]_0^{\Lambda} \propto \Lambda^2
\ee
where $\Lambda$ is the cutoff scale of the theory, where new interactions become effective - e.g.
GUT scale of $10^{16}$ GeV or Plank scale of $10^{19}$ for GUT or gravitational interaction. There are similar quadratically divergent contributions from fermion and gauge boson loops which all tend to push the scalar mass to this cutoff scale $\Lambda$. So the question is how to control the Higgs scalar mass in the desired range of $\sim 10^2$ GeV, which is many orders of magnitude smaller than the cutoff scale $\Lambda$.

   It may be noted here that there are similar divergent loop corrections for gauge boson masses, which cancel out due to gauge symmetry. Similarly for fermion masses they cancel out due to chiral symmetry. There is no such protecting symmetry for the scalar mass. Of course, this feature was used in the last section to give masses to gauge bosons via the Higgs mechanism without breaking gauge symmetry. The hierarchy problem encountered here is the down side of the same coin. The simplest and most attractive solution is to invoke a protecting symmetry - i.e. supersymmetry (SUSY), which is a symmetry between fermions and bosons [17]. As per SUSY all the SM fermions have bosonic superpartners and vice versa. They are listed in Table 4 along with their spins, where the superparticles are indicated by tilde. The superpartners of quarks and leptons are called squarks and sleptons, while those of gauge and Higgs bosons are called photino, gluino, wino, zino and higgsino.

\begin{table}
	   \centering
	   \begin{tabular}{|c|c|c|}\hline
		   quarks \& leptons $\quad$ S $\quad$ & Gauge boson $\quad$ S $\quad$ & Higgs $\quad$ S $\quad$  \\\hline
		  $\,\, q,l$ $\quad$ $\quad$ $\quad$ $\quad$ 1/2 & $\gamma,g,W,Z$ $\quad$  $\quad$1
		   & $h$ $\quad$  $\quad$ 0  \\
           $\tilde q, \tilde l$ $\quad$ $\quad$ $\quad$  $\quad$ 0 & $ \tilde \gamma, \tilde g, \tilde W, \tilde Z$ $\quad$ 1/2
		   & $\,\,$ $\tilde h$ $\quad$ $\quad$ 1/2 
		   \\\hline
	   \end{tabular}
	   \caption{List of SM particles and their Superpartners shown along with their spins.}
	   \label{tab4}
   \end{table}


SUSY ensures cancellation of quadratically divergent contributions between loops of SM particles and their superpartners, e.g. between the Higgs loop of Fig 5 and the corresponding higgsino loop. For the cancellation to occur to the desired accuracy of ~$10^2$ GeV, the mass difference between the superpartners must be restricted to this range. Thus one expects a host of superparticles in the mass range of a few hundred GeV, which can be discovered at the LHC.

   The superparticles are distinguished from the SM particles by a multiplicatively conserved quantity called R parity, i.e.
\be
  R=(-1)^{3 B +L-2 S}
\ee
where B and L represent baryon and lepton numbers. One can easily check from the list shown in Table 4 that all the SM particle have R = +1, while all the superparticles have R = -1. The R parity conservation implies that 1) superparticles are produced in pairs, and 2) each of them decays into the lightest superparticle (LSP), which is stable.

   In most SUSY models the LSP ($\chi$) is an admixture of photino and zino. Like neutrino
   it interacts weakly with matter. Matter is made of quarks and electrons. And the LSP
   interacts with them with the electroweak coupling strength via superparticle exchange,
   while neutrino interacts with similar coupling strength via W boson exchange. Since the
   exchange particles also have similar masses, the two interactions are equally weak. So
   the LSP is expected to leave the detector without a trace like the neutrino. This leads
   to the canonical missing-pT signature for superparticle production. One expects to see
   pair production of strongly interacting superparticles like squarks and gluino at LHC,
\be
   pp \rightarrow \tilde q \tilde q, \tilde g \tilde g, \tilde q \tilde g
\ee
followed by their decays into the LSP,
\be
   \tilde q \rightarrow q \chi, \quad \tilde g\rightarrow \bar q q \chi                                                                                                       \ee
This leads to the vintage jets plus missing-$p_T$ signature for superparticle search at $\bar p p$ and $p p$ colliders [18]. Of course with increasing squark and gluino masses one needs to take into account their cascade decay into the LSP via wino and higgsino states. With this modification the above signature is used now as the most robust signature for superparticle search at LHC [19]. Using this signature one can search for squarks and gluinos at LHC up to a mass range of 2-3 TeV.

   It should be added here that the LSP is the leading candidate for the cosmic dark matter (DM), (which is more appropriately called invisible matter, as it does not emit or absorb EM radiation). By all account, its contribution to the matter density of the universe is about five times as large as that of the standard baryonic matter [20]. To complete the history of the universe shown in Fig 4, at the time of the electroweak phase transition or a little earlier, when the temperature of the universe was a few hundred GeV, it was teeming with a host of superparticles in this mass range along with the SM particles. The superparticles readily decayed into the LSP (DM), which then pair annihilated into the SM particles, i.e.
\be
  \chi \chi \rightarrow \bar q q, l^+ l^-, W^+ W^- \cdots
\ee
In the process the DM density came down to a critical level, after which the annihilation rate could not cope up with the Hubble expansion, i.e. they were expanded away before they could meet one another for annihilation. This is called the freeze-out point. The total DM content of the universe remained frozen ever since. The freeze-out temperature is about 1/20th of the DM mass, i.e. several GeV, when the age of the universe was a fraction of a microsecond. The DM particles are supposed to have played a crucial role in structure formation. Since they were immune to EM interaction, they started experiencing the effect of gravitational perturbation from this very early stage. This led to the formation of clumps of DM (proto galaxies), to which the ordinary baryonic matter was attracted after its decoupling from EM radiation. All quantitative models of structure formation ascribe such a preeminent role to DM.

\section{Concluding Remarks }

Let me conclude with a few remarks, highlighting the main points of LHC physics.
\begin{itemize}

\item	The above mentioned Higgs and superparticles represent the simplest set of missing pieces, required to complete the picture of particle physics a la MSSM.
\item	LHC offers a fairly comprehensive search for Higgs boson(s) up to the expected mass limit of about 1 TeV and for strongly interacting superparticles up to the corresponding mass limit of 2-3 TeV.
\item	It will either discover these particles and complete the picture a la MSSM, or else provide valuable clues for an alternative picture. It should be mentioned here that the case for Higgs discovery at LHC is more compelling than that for SUSY. There are few credible alternatives to Higgs boson(s) in the sub-TeV range. On the other hand, there are several (but less simple) alternatives to SUSY like, little Higgs, extra dimension and extended technicolour models. Practically all of them predict new particles at the sub-TeV range, which can be discovered at LHC.
\item	The lightest superparticle is the leading candidate for the cosmic dark matter, whose contribution to the
    matter density of the universe is five times larger than that of the ordinary baryonic matter.
\end{itemize}

In summary, one of the greatest achievements of modern physics is the unification of the microcosm with the macrocosm. It follows from the basic principles of uncertainty and mass energy equivalence, that when we probe deep inside the sub-atomic space, we come across states of very high energy and mass, which abounded in the early history of the universe. Recreating these particles in the laboratory is like recreating the dinosaurs a la Jurassic Park, but much more significant, because they help us trace back the history of the universe to a very early stage. Recreating the quarks, gluons and their phase transition along with the heavy particles like W, Z and top help us to trace back the history of the universe to the first few picoseconds of its creation. Discovery of Higgs boson(s) at LHC will throw light on the nature phase transition the universe went through during those first few picoseconds; and the discovery of SUSY will throw light on the nature of dark matter that abounds throughout the universe today as a relic of that early history. This answers the question posed in the title.

  \section{Acknowledgements} This work was supported in part by the DAE Raja Ramanna Fellowship.


\begin{thebibliography}{99}


\bibitem{[1]} C. Cowan, F. Reines, F. Harrison, E. Anderson and F. Hayes, Science 124, 103
      (1956).
\bibitem{[2]} G. Danby, J. Gaillard, K. Goulianos, L. Lederman, N. Mistry, M. Schwartz and J.
      Steinberger, Phys. Rev. Lett. 9, 36 (1962).
\bibitem{[3]} C. V. Acher et al., Phys. Lett. 18, 196 (1965).
\bibitem{[4]} DONUT Collaboration: K. Kodama et al., Phys. Lett. B504, 218 (2000).
\bibitem{[5]} B. Richter, Rev. Mod. Phys. 49, 251 (1977).
\bibitem{[6]} M. L. Perl, Nature 275, 273 (1978).
\bibitem{[7]} S. C. C. Ting, Rev. Mod. Phys. 49, 235 (1977).
\bibitem{[8]} L. M. Lederman, Sci. Am. 239 ( No 4), 72 (1978).
\bibitem{[9]} TASSO Collaboration: R. Brandelik et al., Phys. Lett. B86, 243 (1979).
\bibitem{[10]} C. Rubbia, Rev. Mod. Phys. 57, 699 (1985).
\bibitem{[11]} CDF Collaboration: F. Abe et al., Phys. Rev. Lett. 74, 2626 (1995);
        D0 Collaboration: S. Abachi et al., Phys. Rev. Lett. 74, 2632 (1995).
\bibitem{[12]} R. M. Godbole, S. Pakvasa and D. P. Roy, Phys. Rev. Lett. 50, 1539 (1983);
       S. Gupta and D. P. Roy, Z. Phys. C 39, 417 (1988);
       H. Baer, V. Barger and R. J. N. Phillips, Phys. Rev. D39, 3310 (1989).
\bibitem{[13]} P. W. Higgs, Phys. Rev. Lett. 13, 508 (1964);
       F. Englert and R. Brout, Phys. Rev. Lett. 13, 321 (1964).
\bibitem{[14]} For a pedagogical review see, Higgs Hunters' Guide by J. F. Gunion, H. Haber,
       G. L. Kane and S. Dawson, Addison-Wesley, Reading, MA (1990).
\bibitem{[15]} B. K. Bullock, K. Hagiwara and A. D. Martin, Phys. Rev. Lett. 67, 3055 (1991);
       Nucl. Phys. B395 (1993); D. P. Roy, Phys. Lett. B277, 183 (1992); Phys. Lett. B349,
       607 (1999); S. Raychaudhuri and D. P. Roy, Phys. Rev. D53, 4902 (1996).
\bibitem{[16]} A. Djouadi, Phys. Rept. 457, 1 (2008); Phys. Rept. 459, 1 (2008).
\bibitem{[17]} For review see, Perspectives in Supersymmetry, ed. G. L. Kane, World Scientific
        (1998); Theory and Phenomenology of Sparticles by M. Drees, R. M. Godbole and
        P. Roy, World Scientific (2004); Weak Scale Supersymmetry: From superfields to
        scattering events by H. Baer and X. Tata, Cambridge Univ. Press (2006).
\bibitem{[18]} G. L. Kane and J. P. Leville, Phys. Lett. B112, 227 (1982);
        P. R. Harrison and C. H. Llewellyn-Smith, Nucl. Phys. B213, 223 (1983);
        E. Reya and D. P. Roy, Phys. Lett. B141, 442 (1984); Phys. Rev. D32, 645 (1985).
\bibitem{[19]} H. Baer, V. Barger, A. Lessa and X. Tata, JHEP 0909:063 (2009).
\bibitem{[20]} For a review see, G. Bertone, D. Hooper and J. Silk, Phys. Rept. 405, 279 (2005).





\end{thebibliography}
\end{document}